\documentclass[tex,twocolumn,epjc3,tightenlines]{svjour3}
\RequirePackage[T1]{fontenc}
\RequirePackage{graphicx}
\RequirePackage{mathptmx}      % use Times fonts if available on your TeX system
\RequirePackage[numbers,sort&compress]{natbib}
\RequirePackage[colorlinks,citecolor=blue,urlcolor=blue,linkcolor=blue]{hyperref}
\RequirePackage[T1]{fontenc}
\usepackage{fullpage}
\usepackage{slashed}
\usepackage{mathtools}
\usepackage{graphics,bm}
\usepackage{epsfig}
\usepackage{graphicx}
\usepackage{amsmath}
\usepackage{amssymb}
\usepackage{bbold}
\usepackage{wrapfig}
\usepackage{color}
\usepackage{hyperref}
\usepackage{mathtools}
\usepackage{cancel}
\usepackage{multirow}
\usepackage{kantlipsum,widetext}
\usepackage{amsmath,amssymb}
\usepackage{graphics,bm}
\usepackage{graphicx}
\usepackage{bbold}
\usepackage{slashed}
\usepackage{feynmf}
\usepackage{wrapfig}
\usepackage{hyperref}

\usepackage{comment} %usepackage to comment out big sections, added by Martin

    \usepackage[vertfit]{breakurl} 

\newcommand{\bqa}{\begin{eqnarray}}
\newcommand{\eqa}{\end{eqnarray}}

\DeclareMathOperator{\Tr}{Tr}

\begin{document}
\title{Quark condensates and magnetization
in  chiral perturbation theory in a uniform magnetic field}

\author{Prabal Adhikari\thanksref{addr1} \and Jens O. Andersen\thanksref{addr2}}
\thankstext{e1}{e-mail: adhika1@stolaf.edu}
\thankstext{e2}{e-mail:andersen@tf.phys.ntnu.no}
\institute{
  Saint Olaf College, Faculty of Natural Sciences and Mathematics, Physics Department, 1520 Saint Olaf Avenue,
  Northfield, MN 55057, United States 
\label{addr1}
          \and
  Department of Physics, %Faculty of Natural Sciences,
Norwegian University of Science and Technology, H{\o}gskoleringen 5,
N-7491 Trondheim, Norway
\label{addr2}
}
\date{\today}

\maketitle

\begin{abstract}
We reconsider the problem of calculating the vacuum free energy (density)
of QCD and
the shift of the quark condensates in the presence of a uniform background magnetic field
using two-and-three-flavor chiral perturbation theory ($\chi$PT).
Using the free energy, we calculate the degenerate, light quark condensates in the two-flavor and the up, down and strange quark condensates in the three-flavor case.
We also use the vacuum free energy to calculate the (renormalized) magnetization of the QCD vacuum, which shows that it is paramagnetic. We find that the three-flavor light-quark condensates and (renormalized) magnetization are improvements on the two-flavor results. We also find that the average light quark condensate is in agreement with the lattice up to $eB=0.2 {\rm\ GeV^{2}}$, and the (renormalized) magnetization is in agreement up to $eB=0.3 {\rm\ GeV^{2}}$, while three-flavor $\chi$PT, which gives a non-zero shift in the difference between the light quark condensates, unlike two-flavor $\chi$PT, underestimates the difference compared to lattice QCD.
\end{abstract}

%\tableofcontents

\section{Introduction}
%QCD in extreme conditions, i.e. at high temperature, high densities, and in
%strong magnetic fields, has received a lot of attention since the first
%phase diagram in the $\mu_B$--$T$ plane appeared more than four decades ago.

QCD in a strong magnetic field has received a lot of attention
for many years due to various applications in high-energy physics and
astrophysics; In non-central heavy ion collisions, where strong
magnetic field of the order $eB\sim m_{\pi}^2$ are generated, in the early universe, and in neutron stars. 
In addition, it is of fundamental interest to 
study how the QCD vacuum responds to external fields.
The QCD vacuum is characterized by the spontaneous breakdown of chiral symmetry and the appearance of $n_f^2-1$ pseudo-Goldstone bosons.
The inclusion of a constant magnetic field increases the magnitude of the quark condensate in the QCD vacuum, which is an example of magnetic catalysis, first discussed in Refs.~\cite{lemmer,sugu,klim1,klim2,klim3,gus1,gus2}. Magnetic catalysis, in general, is the the enhancement by an external magnetic field of a condensate, 
which can either be the expectation value of a fundamental field (e.g. Higgs type) or a composite field (e.g quark condensate).
If the condensate vanishes for zero magnetic field, the magnetic field
induces a condensate and the breakdown of a symmetry for the $B=0$ vacuum.
In this case, the phenomenon is referred to as dynamical symmetry breaking by a magnetic field, which was first pointed out in Ref.~\cite{klim1}. 
Since then, a large number of papers have studied the symmetry breaking effects of an external magnetic field, see~\cite{shovkovyrev, schmittrev}
for an overview. 
The physics behind chiral symmetry breaking in fermionic theories in strong
magnetic fields was explained in Ref.~\cite{gus1}.
The spectrum associated with a spin-${1\over2}$ particle in a uniform magnetic field along the $z$-axis is $E_k=\sqrt{p_z^2+m^{2}+(2k+1-2s_{z})|qB|}$, where $m$ is the mass, $p_z$ is the momentum
in the direction of the magnetic field, $k=0,1,2...$, and $s_{z}=\pm{1\over2}$.
The lowest Landau level has $k=0$ and together with $s_{z}={1\over2}$, we obtain
the lowest energy level, which happens to be independent of the magnetic field,
$E_0=\sqrt{p_z^2+m^2}$.
If the magnetic field is strong, $|qB|\gg m^2$, the
excited states are widely separated from the lowest Landau level (with $s_{z}={1\over2}$)
and from effective field theory arguments, one expects that the higher Landau levels decouple from the dynamics at low energies. Since the dispersion relation associated with the ground state,
$E_{k=0}$, is that of a particle of mass $m$ and momentum $p_z$, the system
is effectively reduced from three to one spatial dimensions.
The dimensional reduction can also be inferred from the 
fermion propagator in a magnetic field by isolating the $k=0$ contribution.
It is worth pointing out that there is no dimensional reduction in the
bosonic case since the spectrum in this case has a large $B$-dependent gap. For instance, assuming $s_{z}=0$, the dispersion relation is $E_k=\sqrt{p_z^2+m^2+(2k+1)|qB|}$.

The gap equation in the NJL model 
that determines the fermionic mass gap $M$ has an interesting structure~\cite{gus2}. 
In the absence of a magnetic field, there is the trivial solution, $M=0$,
while for cutoffs larger than a critical cutoff, there is an additional
nontrivial solution. For nonzero $B$, there is only the nontrivial solution,
leading to dynamical symmetry breaking as noted above.
Moreover, the solution to the gap equation has the same dependence on
the coupling constant $G$ as the gap equation in superconductivity with zero magnetic field.

Further insight into the enhancement of the quark condensate has been provided
by lattice simulations of QCD in a uniform magnetic field~\cite{cherno,balib,delia,bruck}.
The first simulations of QCD in a magnetic background were carried out for two colors in the 
quenched approximation, i.e. with a quark determinant set to one~\cite{cherno}.
In this approximation, the magnitude of the quark condensate increases with the magnetic field and the enhancement was explained in terms of the Banks-Casher relation~\cite{banks}.
Using the path-integral representation of the partition function, one can find an expression for the quark condensate. The magnetic field $B$ appears 
both in the quark determinant and the operator itself (the quark propagator). In this way, one can disentangle the contributions from
the operator (called the valence effect) and the determinant (called the sea effect) and at $T=0$, the latter is also contributing to the enhancement of the quark condensate.

In the present paper, we reconsider QCD in a strong magnetic background
$B$ using chiral perturbation theory~\cite{smilga, agasian,werbos1,werbos2,hofmann}.
Since $\chi$PT is a low-energy effective theory of QCD based on the spontaneous symmetry breaking of the QCD vacuum that gives rise to its pseudo-Goldstone degrees of freedom~\cite{gasser1,gasser2}, the results are expected to be model independent for weak magnetic fields, i.e. $eB\ll \Lambda^{2}_{\rm had}$, where $\Lambda_{\rm had}\sim 4\pi f_{\pi}$. As such it is of interest to compare predictions of $\chi$PT with those of lattice QCD~\cite{qqlat,maglat}.

The article is organized as follows. In the section \ref{2fsection}, we 
briefly discuss the two-flavor $\chi$PT Lagrangian and calculate the
one-loop vacuum free energy, the light quark condensates and the renormalized magnetization.  
In section~\ref{3f}, we calculate the vacuum free energy, the
light and the strange condensates, and the renormalized magnetization in three-flavor $\chi$PT, also at one loop.
In section~\ref{num}, we discuss our results and compare them to lattice simulations.

\section{Two-flavor $\chi$PT}
\label{2fsection}
\subsection{Lagrangian}
As mentioned in the introduction, $\chi$PT is a low-energy theory of QCD
which is based only on its global symmetries and the low-energy 
degrees of freedom. 
In two-flavor QCD, the degrees of freedom are the
three pions which pseudo-Goldstone bosons. In the chiral limit, the
symmetry group of the QCD Lagrangian is $SU(2)_L\times SU(2)_R$. For nonzero quark masses, the
symmetry group is $SU(2)_V$. The terms in the chiral Lagrangian 
can be organized in a low-energy expansion,
\bqa
{\cal L}&=&{\cal L}_2+{\cal L}_4+{\cal L}_6+..\;,
\eqa
where the subscript indicates the order of the Lagrangian in the chiral expansion. Since the calculation in this paper is at $\mathcal{O}(p^{4})$, we will consider the one-loop contributions from $\mathcal{L}_{2}$, which are formally $\mathcal{O}(p^{4})$ and the tree-level counterterms from $\mathcal{L}_{4}$ of the same order.
Since we consider $\chi$PT in an uniform external magnetic field, we omit the terms including virtual photons~\cite{urech}.
The ${\cal O}(p^2)$ Minkowski space Lagrangian is then given by 
\bqa\nonumber
\mathcal{L}_{2}&=&-{1\over4}F_{\mu\nu}F^{\mu\nu}+
\frac{1}{4}f^2{\rm Tr}\left [ \nabla_{\mu} \Sigma \nabla^{\mu} \Sigma^{\dagger}\right ]
\\ 
&&+\frac{1}{4}f^2 {\rm Tr}\left [\chi\Sigma^{\dagger}+\Sigma\chi^{\dagger}\right ]\;,
\label{lag0}
\eqa
where the first term is the contribution from the external magnetic field and $f$ is the bare pion constant. 
\bqa
\chi&=&2B_0M\;,\\ %\nonumber
M&=&{\rm diag}(m_u,m_d)\;.
%={1\over2}(m_u+m_d)\mathbb{1}\\ &&
%+{1\over2}(m_u-m_d)\tau_3\;.
\eqa
In the remainder of this paper, we work in the isospin
limit, i.e. $m_u=m_d$. However, for calculating the quark condensates
it is convenient to distinguish between light quark masses and only
set $m_u=m_d$ at the very end.
The covariant derivatives are defined as
\bqa
\nabla_{\mu} \Sigma&\equiv\partial_{\mu}\Sigma-i\left [v_{\mu},\Sigma \right]-i\{a_{\mu},\Sigma\}\;,\\ 
(\nabla_{\mu} \Sigma)^{\dagger}&=\partial_{\mu}\Sigma^{\dagger}-i [v_{\mu},\Sigma^{\dagger} ]+i\{a_{\mu},\Sigma^{\dagger}\}\;,
\eqa
where $v_{\mu}$ and $a_{\mu}$ are external fields.
Finally, using the exponential form for $\Sigma$,
\bqa
\Sigma=e^{i{\tau_a\phi_a\over f}}\;,
\eqa
where $\phi_a$ are the pion fields parametrizing the Goldstone manifold and $\tau_a$ are the Pauli matrices.

The leading-order Lagrangian is expanded
to second order in the pion fields,
\bqa
{\cal L}_2^{\rm static}&=&f^2B_0(m_u+m_d)\;,
\\
\nonumber
{\cal L}_2^{\rm quadratic}&=&
-{1\over4}F_{\mu}F^{\mu\nu}+
{1\over2}\partial_{\mu}\pi^0\partial^{\mu}\pi^0
\\ && \nonumber
-{1\over2}m_{\pi^0,0}^2(\pi^0)^2
%\\ && %\nonumbera
+D_{\mu}\pi^{+}D^{\mu}\pi^{-}
%+(\partial_{\mu}+ieA_{\mu})\pi^+(\partial^{\mu}-ieA^{\mu})\pi^-
\\ && %\nonumber
-m_{\pi^{\pm},0}^2\pi^+\pi^-\;,
\eqa
where the degenerate, tree-level pion masses are 
\bqa
m_{\pi^0,0}^2&=&B_0(m_u+m_d)\;, \\
m_{\pi^{\pm},0}^2&=&B_0(m_u+m_d)\;.
\eqa
In the following, $m_{\pi,0}$ denotes the tree-level pion mass.
The pion fields are defined as
\bqa
\label{pionfields}
\pi^0&=&\phi_3\;,\hspace{2mm}
\pi^{\pm}={1\over\sqrt{2}}(\phi_1\pm i\phi_2)\;,
\eqa
and the covariant derivatives are defined as
\bqa
\label{cov1}
D_{\mu}\pi^{\pm}&=(\partial_{\mu}\pm ieA_{\mu})\pi^{\pm}\; .
\label{cov2}
\eqa

The next-to-leading order Lagrangian contains 10 linearly independent terms~\cite{gasser1,scherer}, however, 
in a next-to-leading-order calculation, we need only the following terms
\bqa
\nonumber
{\cal L}_4&=&
\frac{1}{16}(l_{3}+l_{4})\left({\rm Tr}\left[\chi\Sigma^{\dagger}+\Sigma\chi^{\dagger}\right]\right)^{2}
\\ \nonumber
&&+{1\over2}(h_1-l_4){\rm Tr}[\chi^{\dagger}\chi]+l_{5}\Tr[F_{\mu\nu}^{R}\Sigma F^{L\mu\nu}\Sigma^{\dagger}]
\\ && 
-{1\over2}(l_5+{4}h_{2})\Tr\left [F_{\mu\nu}^{R}F^{R\mu\nu}+F_{\mu\nu}^{L}F^{L\mu\nu} \right ]\ .
\label{lag}
\eqa
Note in particular that we have omitted terms whose
contributions to the quantities we are computing, vanish in the isospin limit.
The left- and right-handed tensors are defined as
\bqa
F_{\mu\nu}^{R}=\partial_{\mu}r_{\nu}-\partial_{\nu}r_{\mu}-i[r_{\mu},r_{\nu}]\;,\\ F_{\mu\nu}^{L}=\partial_{\mu}\ell_{\nu}-\partial_{\nu}\ell_{\mu}-i[\ell_{\mu},\ell_{\nu}]\;,
\eqa
where $\ell_{\mu}=v_{\mu}-a_{\mu}$ and $r_{\mu}=v_{\mu}+a_{\mu}$ are the left and right-handed fields, respectively.
In the absence of an axial vector potential $a_{\mu}$, $F_{\mu\nu}^{R}=F_{\mu\nu}^{L}=-e\tilde{Q}F_{\mu\nu}$, where $F_{\mu\nu}$ is the electromagnetic tensor and $\tilde{Q}$ is equal to $\frac{\tau_{3}}{2}$ in the two-flavor case and equal to the quark charge matrix $Q={\rm diag}({2\over3},-{1\over3},-{1\over3})$ in the three-flavor case~\cite{gasser1,gasser2,scherer}. This difference arises from the fact that the two-flavor quark charge matrix has a component in the direction of the identity matrix while the three-flavor quark charge matrix does not.
In the following, we choose a constant magnetic field
$B$ pointing in the positive $z$-direction, without loss of generality. In this case, $F_{12}=-F_{21}=-B$ and all other components of $F_{\mu\nu}$ vanish.

Renormalization is carried out by replacing 
the bare couplings $l_i$ and $h_i$ with their renormalized counterparts
$l_i^r$ and $h_i^r$ using the relations
\bqa
\label{bare1}
l_i&=&l_i^r(\Lambda)-
{\gamma_i\Lambda^{-2\epsilon}\over2(4\pi)^2}
\left[{1\over\epsilon}+1\right]\;,
\\
h_i&=&h_i^r(\Lambda)-
{\delta_i\Lambda^{-2\epsilon}\over2(4\pi)^2}
\left[{1\over\epsilon}+1\right]\;,
\label{bare2}
\eqa
%where the superscript $r$ indicates that the coupling is renormalized.
where the necessary coefficients are
\bqa
%\gamma_{1}&=\frac{1}{3},\ %\gamma_{2}=\frac{2}{3},\ 
\gamma_{3}&=&-{1\over2}\;,  \gamma_{4}=2\;, % \\
%\gamma_{5}&=-\frac{1}{6},\ 
%\gamma_{6}=-\frac{1}{3},\ 
%\gamma_{7}=0,\\
\delta_{1}=2,\ \delta_{2}=\frac{1}{12}\;.
%\delta_{3}=0\ .
\eqa
Since the bare couplings are independent of the renormalization scale,
differentiation of Eqs.~(\ref{bare1})--~(\ref{bare2}) with respect to
$\Lambda$ yields the renormalization group equations
for $\epsilon=0$,
\bqa
\label{rgrun}
\Lambda{d\over d\Lambda}l_i^r&=&
-\frac{\gamma_i}{(4\pi)^2}\, %(1+\epsilon)\;,
\\  \Lambda{d\over d\Lambda}h_i^r
&=&-\frac{\delta_i}{(4\pi)^2}\;.%(1+\epsilon)\;.
\label{rgrun2}
\eqa
The low-energy constants $\bar{l}_i$ and $\bar{h}_1$ are defined via
the solutions to the renormalization group equations (\ref{rgrun})--(\ref{rgrun2}) 
  \bqa
  \label{lr}
l_i^r(\Lambda)&=&{\gamma_i\over2(4\pi)^2}\left[\bar{l}_i+\log{M^2\over\Lambda^2}
  \right]\;,\\
h_i^r(\Lambda)&=&{\delta_i\over2(4\pi)^2}\left[\bar{h}_i+\log{M^2\over\Lambda^2}
\right]\;,
\label{hr}
\eqa
where $M$ is a reference scale.
Up to a prefactor, the low-energy constants are simply the running couplings evaluated at the scale $M^2$.
If $\gamma_i=0$ or $\delta_i=0$, 
the expressions are not valid and the couplings 
do not run.

\subsection{One-loop free energy density}
The tree-level free energy $V_0$ is given by the sum of
the negative of the static Lagrangian ${\cal L}_2^{\rm static}$
and the negative of the magnetic field contribution from ${\cal L}_2^{\rm quadratic}$
\bqa
V_0&=&{1\over2}B^2
-f^2m_{\pi,0}^2\;.
\label{stat0}
\eqa
The  counterterms from ${\cal L}_4^{\rm}$ are
\bqa
V_1^{\rm ct}&=&
%-{1\over2}(h_1-l_4-h_3)B_0^2(m_u-m_d)^2+
%-4(l_3+h_1)B_0^2m^2%-{1\over2}h_1B_0^2(m_u+m_d)^2
-(l_3+h_1)m_{\pi,0}^4
+4h_2(eB)^2\;.
\label{stat1}
\eqa
%\pagebreak
The one-loop contribution to the free energy density is
\bqa\nonumber
V_1&=&\log\det[-D_{\mu}D^{\mu}+m_{\pi,0}^2]
\\ \nonumber
&=&{eB\over2\pi}\sum_{k=0}^{\infty}\int_{p_{\parallel}}
\log[p_{\parallel}^2+eB(2k+1)+m_{\pi,0}^2]\\ \nonumber
%&=&{\color{red}-}\int_0^{\infty}{ds\over s}{\rm %Tr}\,e^{-s[-D_{\mu}D^{\mu}+m_{\pi,0}^2]}\\
&=&-
{eB\over2\pi}\sum_{k=0}^{\infty}\int_{p_{\parallel}}
\int_0^{\infty}{ds\over s}
e^{-[p_{\parallel}^2+(2k+1)eB+m_{\pi,0}^2]s}\;,
\eqa
where $k=0,1,2...$ are the Landau levels, $p_{\parallel}^2=p_0^2+p_z^2$ 
and the integral is defined in \ref{app:integrals}.
Summing over the Landau levels and 
integrating over $p_{\parallel}$ gives
\bqa
V_1=-\frac{\mu^{2\epsilon}}{(4\pi)^{2}}\int_0^{\infty}{ds\over s^{3-\epsilon}}e^{-m_{\pi,0}^2 s}{eBs\over\sinh(eBs)}\; ,
\label{pot1}
\eqa
where $\mu=\sqrt{e^{\gamma_{E}}\Lambda^{2}}$
and $\Lambda$ is the renormalization scale associated with the
$\overline{\rm MS}$-scheme. 
This integral is defined as $I_1^B(m_{\pi,0}^2)$ and 
can be evaluated in dimensional regularization, see  \ref{app:integrals} for the relevant result. Since $s$ has negative mass dimensions, the ultraviolet behavior of Eq.~(\ref{pot1})
is determined by small $s$, while the infrared behavior is determined by large $s$.
\begin{widetext}
\noindent
Expanding the integral to $\mathcal{O}(\epsilon^0)$, we obtain
\bqa
V_1=&&\nonumber
-{m_{\pi,0}^{4}\over2(4\pi)^2}\left[{1\over\epsilon}+{3\over2}+
\log{\Lambda^2\over m_{\pi,0}^2}
\right]+{(eB)^2\over6(4\pi)^2}\left[
{1\over\epsilon}+\log{\Lambda^2\over m_{\pi,0}^2}
\right]+{4(eB)^2\over(4\pi)^2}\Bigg[\zeta^{(1,0)}(-1,\tfrac{1}{2}+x)
+{1\over4}x^2\Bigg.\\
&&\Bigg.-{1\over2}x^2\log x+{1\over24}\log x+{1\over24}\Bigg]
\;,
\label{swing}
\eqa
where we have defined the dimensionless quantity $x={m_{\pi,0}^2\over2eB}$ and $\zeta(a,x)$ is the Hurwitz zeta-function with derivatives with respect to the two variables indicated using numbers in the parenthesis of the superscript.
%The integrals $I_n^B(m^2)$ are defined in Eq.~(\ref{indef}).
Eq.~(\ref{swing}) is the famous (unrenormalized)
result obtained by Schwinger~\cite{svinger}.
The result is divergent in the ultraviolet, i.e. for small $s$, and it therefore requires renormalization, which will be done below. The result for $V_1$ is well behaved in the chiral limit, which 
follows directly from evaluating Eq.~(\ref{pot1}), noting
that it converges for large $s$. 
%Alternatively,
%the infrared divergence quadratic in the external magnetic field in the second bracket is cancelled by the infrared divergence in the term within the last bracket. 
If we expand $V_1$ in powers of $eB$, we notice that each term is increasingly divergent in the limit $m_{\pi,0}^2\rightarrow 0$.
Each term can be thought of as a one-loop graph of $\mathcal{O}((eB)^{2n})$ with $2n$-propagators and $2n$-field insertions, where $n$ is a positive integer, with the degree of infrared divergence increasing with $n$~\cite{smilga}. While each of these graphs is infrared divergent, the sum is convergent in the chiral limit. 
%This is reminiscent of the resummation
%of infrared divergent graphs that is needed in high-temperature field %theory to obtain a finite result.
The complete one-loop result is the sum of 
Eqs.~(\ref{stat0}), (\ref{stat1}), and~(\ref{swing}).
Renormalization is carried out by replacing the bare couplings with the renormalized ones, which yields
%After renormalization, we obtain the complete one-loop vacuum energy 
\bqa%\nonumber
V_{0+1}&=&{1\over2}B_r^2
%-{1\over2}(h_1^r-l_4^r-h_3^r)B_0^2(m_u-m_d)^2\\ && 
-f^2m_{\pi,0}^2\left\{1+\left[l_3^r+h_1^r
+{3\over4(4\pi)^2}\left(
\log{\Lambda^2\over m_{\pi,0}^2}+{1\over2}\right)\right]{m_{\pi,0}^2\over f^2}
\right\}
%+\cancel{{(eB)^2\over(4\pi)^2}\tilde{I}_1^{B}(m_{\pi,0}^2)}
+\frac{(eB)^{2}}{(4\pi)^{2}}\delta\tilde{I}^{B}_{1}(m_{\pi,0}^2)\;,
\label{v12f}
\eqa
%\noindent
where $\delta\tilde{I}_{1}^{B}(m_{\pi,0}^2)$ is defined in Eq.~(\ref{dI1B}) and the renormalized magnetic field $B_r$ is
\end{widetext}
\bqa\nonumber
B_r^2&=&B^{2}\left\{1+{e^2\over3}\left[24h_2^r+{1\over(4\pi)^2}
\left(\log{\Lambda^2\over m_{\pi,0}^2}-1\right)\right]\right\}\;.
\\ &&
\label{bren}
\eqa
Using the renormalization group equations for the running
couplings, one finds that the results (\ref{v12f}) and~(\ref{bren})
are independent of the scale $\Lambda$.

\subsection{One-loop quark condensate and magnetization}
The quark condensate is denoted by $\langle\bar{q}q\rangle$,
where $q=u,d$. Since we are in the isospin limit, they
coincide and are defined as
\bqa
\langle\bar{q}q\rangle&=&
{\partial V_{0+1}\over\partial m_{q}}\;.
\label{lightdef}
\eqa
We separate the condensates into a magnetic field independent and dependent contribution, $\langle\bar{q}q\rangle=
\langle\bar{q}q\rangle_0+\langle\bar{q}q\rangle_B$, where the
first term is the vacuum value of the condensate
and the second is the shift of the condensate due the magnetic field.
The quark condensate is
\begin{widetext}
\bqa
\langle\bar{u}u\rangle=\langle\bar{d}d\rangle
&=&
-f^2B_0\left\{1+
\left[2l_3^r+2h_1^r+{3\over2(4\pi)^2}
\log{\Lambda^2\over m_{\pi,0}^2}\right]
%\\ \nonumber &&\times 
{m_{\pi,0}^2\over f^2}+\frac{eB}{(4\pi)^{2}f^{2}}
\tilde{I}^{B}_{2}(m_{\pi,0}^{2})
\right\}
\;,
\eqa
where $\tilde{I}_2^B$ is defined in Eq.~(\ref{i2b}). In the chiral limit, $\tilde{I}_2^B(m_{\pi,0}^2)$ reduces
to $\log2$ and therefore
\bqa
\langle\bar{q}q\rangle&=&\langle\bar{q}q\rangle_0
\left[1+{eB\log2\over(4\pi)^2f^2}\right]\;,
\eqa
as first obtained in~\cite{smilga}.
\end{widetext}

Next, we consider the renormalized magnetization of the QCD vacuum, which has been studied analytically in a model-independent setting for large magnetic fields, i.e. $eB\gg \Lambda_{\rm had}$~\cite{werbos3} but not the opposite regime. The renormalized magnetization is the response of the QCD vacuum in the presence of a magnetic field and is defined as~\cite{hrg}
\bqa
M_{r}&=&-{\partial \tilde{V}_{0+1}\over\partial (eB)}\ ,
\eqa
where $\tilde{V}_{0+1}=V_{0+1}-\frac{1}{2}B_{r}^{2}$. The definition is motivated by the physical origin of the renormalized magnetization: in the presence of an external magnetic field, virtual, charged meson pairs that pop out of the vacuum interact with the external magnetic field giving rise to current loops that magnetize the QCD vacuum. Since these vacuum fluctuations decrease with increasing mass, the magnetization vanishes in the limit of large meson masses. With this physical intuition in mind, it is straightforward to show using the one-loop effective potential that the contribution of each pair of charged mesons to the renormalized magnetization is
%\bqa\nonumber
%M_{r}(eB)=&-\frac{8eB}{(4\pi)^{2}}\bigg[\zeta^{(1,0)}
%\left(-1,x+\tfrac{1}{2}\right)\bigg.\\ \nonumber
%&\left.
%-{1\over2}x \zeta^{(1,1)}\left(-1,x+\tfrac{1}{2}\right)+{1\over4}
%x^{2}
%\right.\\
%&\bigg.+\frac{1}{24}\log x+\frac{1}{48}\bigg]\;.
%\eqa
\begin{align}\nonumber
M_{r}(eB)=&-\frac{8eB}{(4\pi)^{2}}\bigg[\zeta^{(1,0)}
\left(-1,x+\tfrac{1}{2}\right)\bigg.\\ \nonumber
&\left.
-{1\over2}x \zeta^{(1,1)}\left(-1,x+\tfrac{1}{2}\right)+{1\over4}
x^{2}
\right.\\
&\bigg.+\frac{1}{24}\log x+\frac{1}{48}\bigg]\;.
\end{align}
While it is not immediately obvious that $M^{r}$ is suppressed at large $m_{\pi,0}$, we note that $M_{r}$ can be written in the Schwinger proper time form,
\begin{equation}
    M_{r}(eB)=\frac{1}{(4\pi)^{2}}\int_{0}^{\infty}{ds\over s^2} e^{-m_{\pi,0}^{2}s}f(eBs)\;,
\end{equation}
where $f(z)$ is a well-behaved function independent of $m_{\pi,0}$: $f(z)=\left(1-\frac{z}{\tanh z}\right)\frac{1}{\sinh z}+\frac{z}{3}$.
In this representation, it is immediately obvious that the renormalized magnetization, $M_{r}\rightarrow 0$, as $m_{\pi,0}\rightarrow \infty$, which is consistent with the physical
picture. Since $M_{r}$ is positive definite (vanishing in the absence of the magnetic field), the QCD vacuum is paramagnetic~\cite{hrg}.

\section{Three-flavor $\chi$PT} 
\label{3f}
In this section we will calculate the vacuum free energy, quark condensates, and the  magnetization for three-flavor $\chi$PT. We will restrict ourselves to a one-loop calculation in the isospin limit.

\subsection{Lagrangian}
In three-flavor $\chi$PT, the low-energy degrees are,  in addition to the pions,  
the charged and neutral kaons and the $\eta$ meson.
The chiral perturbation Lagrangian at $\mathcal{O}(p^{2})$ is still given
by Eq.~(\ref{lag0}), but the mass
matrix is
\bqa
%\chi&=&s+ip=2B_{0}M\;,\\ \nonumber
M&=&\textrm{diag}{(m_{u},m_{d},m_{s})}\;,
%\\ \nonumber
%&=&\tfrac{1}{\sqrt{6}}(m_{u}+m_{d}+m_{s})\mathbb{1}+\tfrac{1}{2}(m_{u}-m_{d})\lambda_{3}
%\\ &&+\tfrac{1}{2\sqrt{3}}(m_{u}+m_{d}-2m_{s})\lambda_{8}\;,
\eqa
where $m_{u}=m_{d}$ in the isospin limit, but we differentiate between the up-quark and down-quark masses since we will calculate the up and down quark condensates separately. We use the exponential form for the SU(3) matrix, $\Sigma$,
\bqa
\Sigma=\exp\left (\tfrac{\lambda_{a}\phi_{a}}{f} \right)\ ,
\eqa
which contains fields associated with the meson octet.

Expanding the leading order Lagrangian to second order in the mesonic fields
gives
\bqa\nonumber
\mathcal{L}_{2}^{\rm static}&=&f^{2}B_{0}\left(m_{u}+m_{d}+m_{s}\right)\;,\\
\nonumber
\mathcal{L}_{2}^{\rm quadratic}&=&
-{1\over4}F_{\mu}F^{\mu\nu}+
\frac{1}{2}\partial_{\mu}\pi^{0}\partial^{\mu}\pi^{0}
-\frac{1}{2}m_{\pi^{0},0}^{2}(\pi^{0})^{2}
\\  \nonumber
 &&
+D_{\mu}\pi^{+}D^{\mu}\pi^{-}
-m_{\pi^{\pm},0}^{2}\pi^{-}\pi^{+}
\\ && \nonumber
+D_{\mu}K^{+}D^{\mu}K^{-}
-m_{K^{\pm},0}^{2}K^{+}K^{-}
\\ && \nonumber
+\partial_{\mu}K^{0}\partial^{\mu}\bar{K}^{0}
-m_{K^{0},0}^{2}K^{0}\bar{K}^{0} \\ 
&&
+\frac{1}{2}\partial_{\mu}\eta\partial^{\mu}\eta
-\frac{1}{2}m_{\eta,0}^{2}\eta^{2}\;.
%+\frac{1}{2}m^{2}_{\pi_{0}\eta}\pi_{0}\eta+\frac{1}{2}m^{2}_{\pi_{0}\eta}\eta\pi_{0}\;.
\eqa
where the tree-level masses are
\bqa
m_{\pi^{\pm},0}^{2}&=&B_{0}(m_{u}+m_{d})\;,\\ 
m_{\pi^{0},0}^{2}&=&B_{0}(m_{u}+m_{d})\;,\\ 
m_{K^{\pm},0}^{2}&=&B_{0}(m_{u}+m_{s})\;,\\ m_{K^{0},0}^{2}&=&B_{0}(m_{d}+m_{s})\;,\\ 
m_{\eta,0}^2&=&{B_0(4m_s+m_{u}+m_{d})\over3}\;.
%m_{\pi_{0}\eta}^{2}&=&\frac{B_{0}(m_{d}-m_{u})}{\sqrt{3}}\;.
\eqa
The pion fields are defined in Eq.~(\ref{pionfields}) and the kaon fields are defined as
\bqa
K^{\pm}&=&{1\over\sqrt{2}}(\phi_4\pm i\phi_5)\;,\\
K^{0}/\bar{K}^0&=&{1\over\sqrt{2}}(\phi_6\pm i\phi_7)\;,\\
\eta&=&\phi_8\;.
\eqa
The covariant derivative of the charged kaons $K^{\pm}$ is defined as for the pions in Eqs.~(\ref{cov1})--(\ref{cov2}). In the following, we write $m_{\pi,0}$ for the tree-level pion 
as in the two-flavor case. There is no mixing between the neutral pion and the eta meson because we are working in the isospin limit.

In order to calculate the one-loop vacuum free energy, we also need the Lagrangian at $\mathcal{O}(p^{4})$ given in Ref.~\cite{gasser2}. The relevant terms are
\bqa\nonumber
\mathcal{L}_{4}&=&
L_{6}\left [\Tr(\chi\Sigma^{\dagger}+\chi^{\dagger}\Sigma)\right ]^{2}
\\  \nonumber
&+&L_{8}\Tr \left[\Sigma \chi^{\dagger}\Sigma \chi^{\dagger}+\chi\Sigma^{\dagger}\chi\Sigma^{\dagger}\right]
\\ \nonumber
&+&L_{10}\Tr \left[\Sigma F^{L}_{\mu\nu}\Sigma^{\dagger}F^{R\mu\nu} \right]
\\ %\nonumber
&+&H_{1}\Tr\left [F_{\mu\nu}^{R}F^{R\mu\nu}+F_{\mu\nu}^{L}F^{L\mu\nu} \right ]
+H_{2}\Tr[\chi^{\dagger}\chi]
\;,
%\\
\eqa
where $L_{i}$ with $i=1,2\dots 10$ and $H_{i}$ with $i=1,2$ are the bare couplings. The bare and renormalized couplings are related as
\bqa
L_{i}&=&L_{i}^{r}(\Lambda)-
{\Gamma_{i}\Lambda^{-2\epsilon}\over2(4\pi)^2}
\left[{1\over\epsilon}+1\right]\;,
\\ 
H_{i}&=&H_{i}^{r}(\Lambda)-
{\Delta_{i}\Lambda^{-2\epsilon}\over2(4\pi)^2}
\left[{1\over\epsilon}+1\right]\;.
\eqa
%\bqa\lambda=-\frac{\Lambda^{-2\epsilon}}{2(4\pi)^{2}}\left (\frac{1}{\epsilon}+1 \right )\eqa
with
\bqa
\Gamma_{6}&=&\frac{11}{144}\;,\
\Gamma_{8}=\frac{5}{48}\;,\ \Gamma_{10}=-\frac{1}{4}\;,\\
\Delta_{1}&=&-\frac{1}{8}\;,\ \Delta_{2}=\frac{5}{24}\;.
\eqa
The renormalization group equations for the running couplings
are now
\bqa
\label{rgrun3}
\Lambda{d\over d\Lambda}L_i^r&=&-
\frac{\Gamma_i}{(4\pi)^2}\;,%(1+\epsilon)\;,
\\  \Lambda{d\over d\Lambda}H_i^r
&=&-\frac{\Delta_i}{(4\pi)^2}\;.%(1+\epsilon)\;.
\label{rgrun4}
\eqa

\subsection{One-loop free energy density}
The one-loop contributions to the vacuum free energy from the charged pions and
the kaons are of the same form as in the two-flavor case. We assume the isospin limit but separate the contributions of the charged and the neutral kaons.
%However, away
%from the isospin limit, there is a small mixing angle between the $\pi^0$ and $\eta$ giving rise to mass eigenstates $\tilde{\pi}_{0}$ and $\tilde{\eta}$, which we discuss in \ref{mixing}.
The tree-level contribution to the vacuum free energy is
\bqa
V_0&=&
{1\over2}B^2
-{1\over2}f^2(m_{\pi,0}^2+m_{K^{\pm},0}^2+m_{K^{0},0}^2)\;,
\eqa
the counterterm contribution is
\bqa\nonumber
V_1^{\rm ct}&=&
-6L_6\left(m_{\pi,0}^4+4m_{K^{\pm},0}^4+4m_{K^{0},0}^4-3m_{\eta,0}^4\right)\\&&\nonumber
%\textcolor{magenta}{-4L_6\left(m_{\pi,0}^2+m_{K^{0},0}^2+m_{K^{\pm},0}^2\right)^{2}}\\
%-16L_6B_0^2(m_u+m_d+m_s)^2
\\ && \nonumber
-(2L_8+H_2)\left({5\over2}m_{\pi,0}^4-2m_{K^{\pm},0}^4-2m_{K^{0},0}^4
+{9\over2}m_{\eta,0}^4\right)
%B_0^2(m_u^2+m_d^2+m_s^2)
\\ &&
%-{2\over3}(L_{10}+2H_1)F_{\mu\nu}F^{\mu\nu}\;.\\ &&
{
-{4\over3}(L_{10}+2H_1)(eB)^2}\;,
\eqa
and the one-loop contribution is
\bqa\nonumber
V_1&=&{1\over2}I_1(m^2_{\pi,0})+I_1^B(m^2_{\pi,0})
+I_1^B(m^2_{K^{\pm},0})
\\ &&
+I_1(m^2_{K^{0},0})+{1\over2}I_1(m^2_{\eta,0})\;,
\eqa
where the integrals $I_n(m^2)$ and $I_n^B(m^2)$ are defined in \ref{app:integrals}.
Renormalization is carried out as in the two-flavor case; the
complete NLO result for the vacuum free energy is
\begin{widetext}
\bqa\nonumber
V_{0+1}&=&
{1\over2}B_r^{2}
-{1\over2}f^2(m_{\pi,0}^2+m_{K^{\pm},0}^2+m_{K^{0},0}^2)
-\left[6L^{r}_{6}+5L_8^r+{5\over2}H_2^r+{3\over4(4\pi)^2}\left(\log{\Lambda^2\over m_{\pi,0}^2}+{1\over2}\right)\right]{m_{\pi,0}^4}
%-{1\over2}f^2(m_{\pi^{\pm},0}^2+m_{K^{\pm},0}^2+m_{K^0,0}^2)
%-4L_6(m_{\pi,0}^2+2m_{K,0}^2)^2
%\\
%\nonumber
%&&
%-(2L_8^r+H_2^r)\left[
%3m_{\pi_{\pm},0}^4+3m_{K^{\pm},0}^4+3m_{K^0,0}^4)
%-2m_{\pi_{\pm},0}^2m_{K^+,0}^2-2m_{\pi_{\pm},0}^2m_{K^0,0}^2-2m_{K^+,0}^2m_{K^0,0}^2
%\right]
%\\  \nonumber
%&&
%-{1\over2}f^2m_{\pi^{\pm},0}^2\left\{1+
%\left[8L_6^r+{1\over(4\pi)^2}\left(\log{\Lambda^2\over m_{\pi^{\pm},0}^2}+{1\over2}\right)
%\right]{m_{\pi^{\pm},0}^2\over f^2}\right\}
%\\  \nonumber
%\nonumber
%&&-\cancel{{1\over2}}f^2m_{K,0}^2\left\{1+
%\left[8L_6^r+{1\over(4\pi)^2}\left(\log{\Lambda^2\over m_{K}^2}+{1\over2}\right)
%\right]{m_{K,0}^2\over f^2}\right\}
%\\ \nonumber
%&&\cancel{-{1\over2}f^2m_{K^0,0}^2\left\{1+
%\left[8L_6^r+{1\over(4\pi)^2}\left(\log{\Lambda^2\over m_{K^0}^2}+{1\over2}\right)
%\right]{m_{K^0,0}^2\over f^2}\right\}}
\\ \nonumber
&-&\left[24L_6^r-4L_8^r-2H_2^r+{1\over2(4\pi)^2}\left(\log{\Lambda^2\over m_{K^{\pm},0}^2}+{1\over2}\right)\right]{m_{K^{\pm},0}^4}
-\left[24L_6^r-4L_8^r-2H_2^r+{1\over2(4\pi)^2}\left(\log{\Lambda^2\over m_{K^0,0}^2}+{1\over2}\right)\right]{m_{K^0,0}^4}
%-{m_{\pi^0,0}^4\over4(4\pi)^2}\left(\log{\Lambda^2\over m_{\pi^0,0}^2}+{1\over2}\right)
\\
&-&\left[-18L_6^r+9L_8^r+{9\over2}H_2^r+{1\over4(4\pi)^2}\left(\log{\Lambda^2\over m_{\eta,0}^2}+{1\over2}\right)\right]{m_{\eta,0}^4}
%\cancel{-{4(eB)^2\over(4\pi)^2}\left[\tilde{I}_1^B(m_{\pi^{\pm},0}^2)+\tilde{I}_1^B(m_{K^{\pm},0}^2)\right]}\\&&
+\frac{(eB)^{2}}{(4\pi)^{2}}\left[\delta\tilde{I}^{B}_{1}(m_{\pi,0}^2)+\delta\tilde{I}^{B}_{1}(m_{K^{\pm},0}^2)\right]
\;,
\label{v1ren}
\eqa
%where $i$ is the meson index, i.e. $i=\pi^{\pm}$, $K^{\pm}$, $K^{0}$, $\bar{K}^{0}$ %$\tilde{\pi}^{0}$ and $\tilde{\eta}$ and
%$c_i=1$ for neutral mesons and $c_i=2$ for charged mesons.
where the renormalized magnetic field is
\bqa%\nonumber
B_r^{2}&=&B^2\left\{1+{e^2\over3}\left[-8(L_{10}^r+2H_1^r)
+{1\over(4\pi)^2}\left(\log{\Lambda^2\over m_{\pi,0}^{2}}+
\log{\Lambda^2\over m_{K^{\pm},0}^{2}}-2\right)\right]\right\}\;.
\eqa
The renormalized vacuum free energy and the renormalized magnetic field are scale independent 
as can be verified by using the renormalization group equations for the couplings. 

\subsection{One-loop quark condensates and magnetization}
%\noindent
%The up-, down- and strange-quark condensates, denoted $\langle\bar{q}q\rangle$ can be written in the form
%\begin{equation}
%\begin{split}
%\langle\bar{q}q\rangle&=\langle\bar{q}q\rangle_{0}+\langle\bar{q}q\rangle_{H}\;,
%\end{split}
%\end{equation}
%where $\langle\bar{q}q\rangle_{0}$ are the $H=0$ condensates where $q=u,d,s$ and $\langle\bar{q}q\rangle_{H}$ is the condensate shift in a uniform magnetic field. 
The light quark condensates are defined in Eq.~(\ref{lightdef}) and the $s$-quark condensate is defined as
\bqa
\langle\bar{s}s\rangle&=&{\partial V\over\partial m_s}\;.
\eqa
This yields
\bqa\nonumber
\langle\bar{u}u\rangle&=&-f^{2}B_{0}
\left\{1+
%\left[
%{1\over(4\pi)^2}\log{\Lambda^2\over m_{\pi^0,0}^2}\right]{m_{\pi^0,0}^2\over2f^2}+
\left[12L_6^r+10L_8^r+5H_2^r+
{3\over2(4\pi)^2}\log{\Lambda^2\over m_{\pi,0}^2}\right]{m_{\pi,0}^2\over f^2}
+\left[48L_6^r-8L_8^r-4H_2^r+{1\over(4\pi)^2}\log{\Lambda^2\over m_{K^{\pm},0}^2}\right]{m_{K^{\pm},0}^2\over f^2}
\right.  \\ 
&&\left.
+\left[-12L_6^r+6L_8^r+3H_2^r+{1\over6(4\pi)^2}\log{\Lambda^2\over m_{\eta,0}^2}\right]{m_{\eta,0}^2\over f^2}
+{eB\over(4\pi)^2f^2}\left[\tilde{I}_2^B(m_{\pi,0}^2)+
\tilde{I}_2^B(m_{K^{\pm},0}^2)\right]
\right\}\;, \\ \nonumber
\langle\bar{d}d\rangle&=&-f^{2}B_{0}
\left\{1+
%\left[+{1\over(4\pi)^2}\log{\Lambda^2\over m_{\pi,0}^2}
%\right]{m_{\pi^0,0}^2\over2f^2}+
\left[12L_6^r+10L_8^r+5H_2^r+{3\over2(4\pi)^2}\log{\Lambda^2\over m_{\pi,0}^2}\right]{m_{\pi,0}^2\over f^2}
%\right.  \\  \nonumber &&\left.
+\left[48L_6^r-8L_8^r-4H_2^r+{1\over(4\pi)^2}\log{\Lambda^2\over m_{K^0,0}^2}\right]{m_{K^0,0}^2\over f^2}
\right.  \\ &&\left.
+\left[-12L_6^r+6L_8^r+3H_2^r+
{1\over6(4\pi)^2}\log{\Lambda^2\over m_{\eta,0}^2}\right]{m_{\eta,0}^2\over f^2}
+{eB\over(4\pi)^2f^2}\tilde{I}_2^B(m_{\pi,0}^2)
\right\}\;, \\ \nonumber
\langle\bar{s}s\rangle&=&-f^{2}B_{0}
\left\{1+
\left[48L_6^r-8L_8^r-4H_2^r+{1\over(4\pi)^2}\log{\Lambda^2\over m_{K^{\pm},0}^2}\right]{m_{K^{\pm},0}^2\over f^2}
+\left[48L_6^r-8L_8^r-4H_2^r+{1\over(4\pi)^2}\log{\Lambda^2\over m_{K^0,0}^2}\right]{m_{K^0,0}^2\over f^2}
%\right. \\  \left. &&
\right.  \\ 
&&\left.
+\left[
-48L_6^r+24L_8^r+12H_2^r
+{2\over3(4\pi)^2}\log{\Lambda^2\over m_{\eta,0}^2}\right]{m_{\eta,0}^2\over f^2}
%+{8B_0m_s\over f^2}\left[2L_8^r+H_2^r\right]
+{eB\over(4\pi)^2f^2}\tilde{I}_2^B(m_{K^{\pm},0}^2)
\right\}\;.
\eqa
\end{widetext}
With no background magnetic field, these expressions reduce to the original
ones in Refs.~\cite{gasser1,gasser2}. Notice also that the
shift in the condensates due the magnetic field satisfy 
$\langle\bar{u}u\rangle_B=\langle\bar{d}d\rangle_B+\langle\bar{s}s\rangle_B$. The origin of this rule is intimately connected to the fact that the shift in the effective potential due to the magnetic background depends on the charged pion and charged kaon masses and that the valence quark and anti-quark in the charged pions (kaons) are up-and-down (up-and-strange).
In contrast to the two-flavor case, 
the difference between the two light quark
condensates receives one-loop corrections,
which are purely $B$-dependent in the isospin limit.
The renormalized magnetization in the three-flavor case has the same form as in the two-flavor case with an additional contribution arising due to the pair of charged kaons. The renormalized magnetization is
%\begin{equation}
%\begin{split}
%M_{r}(eB)&=-\sum_{i}\frac{eB}{6(4\pi)^{2}}\left[1+3\left(\frac{m_{i}^{2}}{eB}\right)^{2}+2\log\frac{m_{i}^{2}}{2eB}\right.\\
%&+48\zeta^{(1,0)}\left(-1,\frac{1}{2}+\frac{m_{i}^{2}}{2eB}\right)\\
%&\left.-\frac{12m_{i}^{2}}{eB} \zeta^{(1,1)}\left(-1,\frac{1}{2}+\frac{m_{i}^{2}}{2eB}\right)\right ]\ ,
%\end{split}
%\end{equation}
%\begin{equation}
%\begin{split}
%M_{r}(eB)&=-\sum_{i}\frac{eB}{6(4\pi)^{2}}\left[1+12x_{i}^{2}+2\log %x_{i}\right.\\
%&+48\zeta^{(1,0)}\left(-1,\tfrac{1}{2}+x_{i}\right)\\
%&\left.-24x_{i} \zeta^{(1,1)}\left(-1,\tfrac{1}{2}+x_{i}\right)\right ]\ ,
%\end{split}
%\end{equation}
\begin{align}\nonumber
M_{r}(eB)=&-{8eB\over(4\pi)^2} \sum_{i=1}^{2}
\bigg[\zeta^{(1,0)}\left(-1,x_{i}+\tfrac{1}{2}\right)\bigg.
\\\nonumber
&\left.-{1\over2}x_{i} \zeta^{(1,1)}\left(-1,x_{i}+\tfrac{1}{2}\right)
+{1\over4}x_{i}^{2}\right.\\
&\bigg.+\frac{1}{24}\log x_{i}+\frac{1}{48}\bigg]\; ,
\end{align}
where $x_{i}=\frac{m_{i}^{2}}{2eB}$ with $m_{1}=m_{\pi,0}$ and $m_2=m_{K^{\pm},0}$.
As such one expects the three-flavor QCD vacuum to exhibit enhanced paramagnetism compared to the two-flavor QCD vacuum.

\section{Results and discussion}
\label{num}
In this section, we compare the results from our one-loop calculation with results from $1+1+1$-flavor lattice QCD in the isospin limit~\cite{qqlat,maglat}. In particular, we focus on the average chiral condensate shift for the light quarks, the shift in the difference between light quark condensate, and the renormalized magnetization. 

\begin{figure}[htb]
\centering  \includegraphics[width=0.45\textwidth]{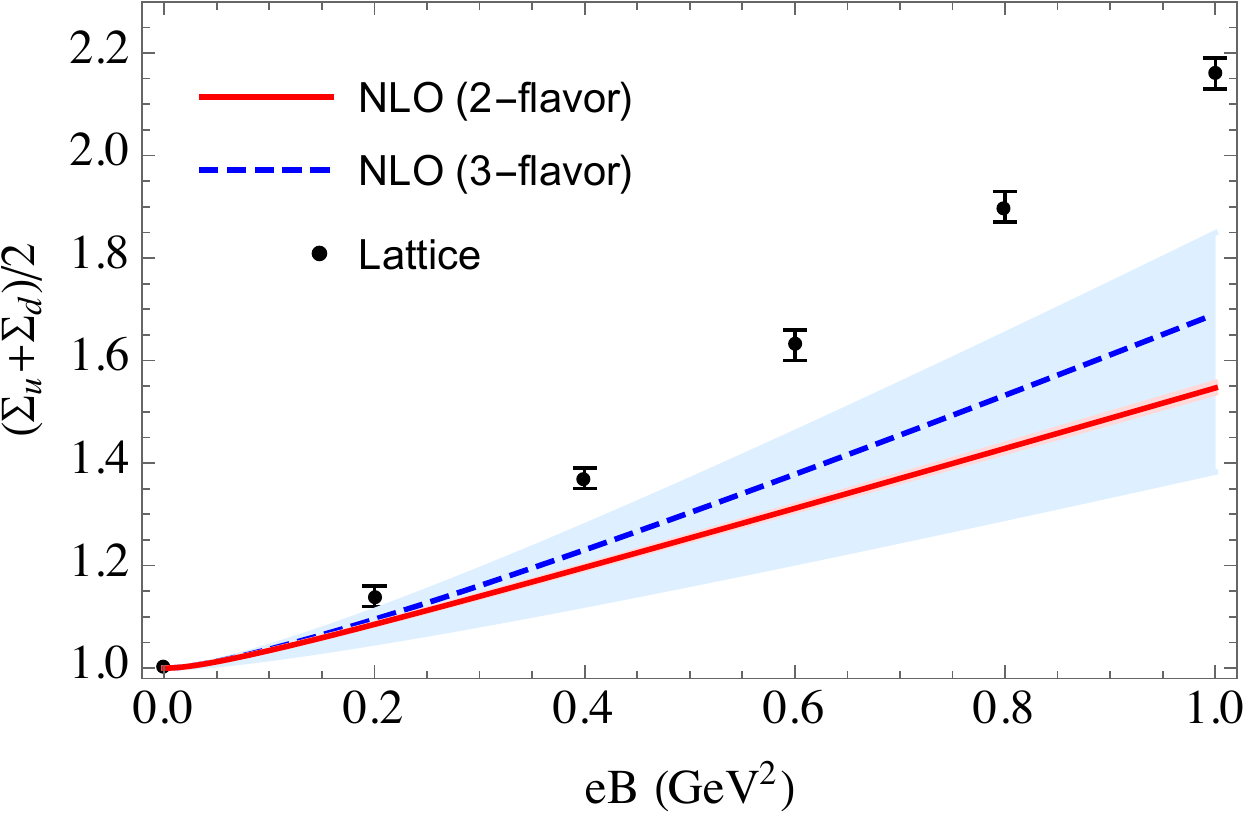}
  \caption{Plot of the average light condensate shift, ${1\over2}(\Sigma_{u}+\Sigma_{d})$, as a function of $eB$ at NLO in two-flavor $\chi$PT (red) and three-flavor $\chi$PT (blue). See main text for details.}
\label{condavg}
\end{figure}

\begin{figure}[htb]
\centering  \includegraphics[width=0.45\textwidth]{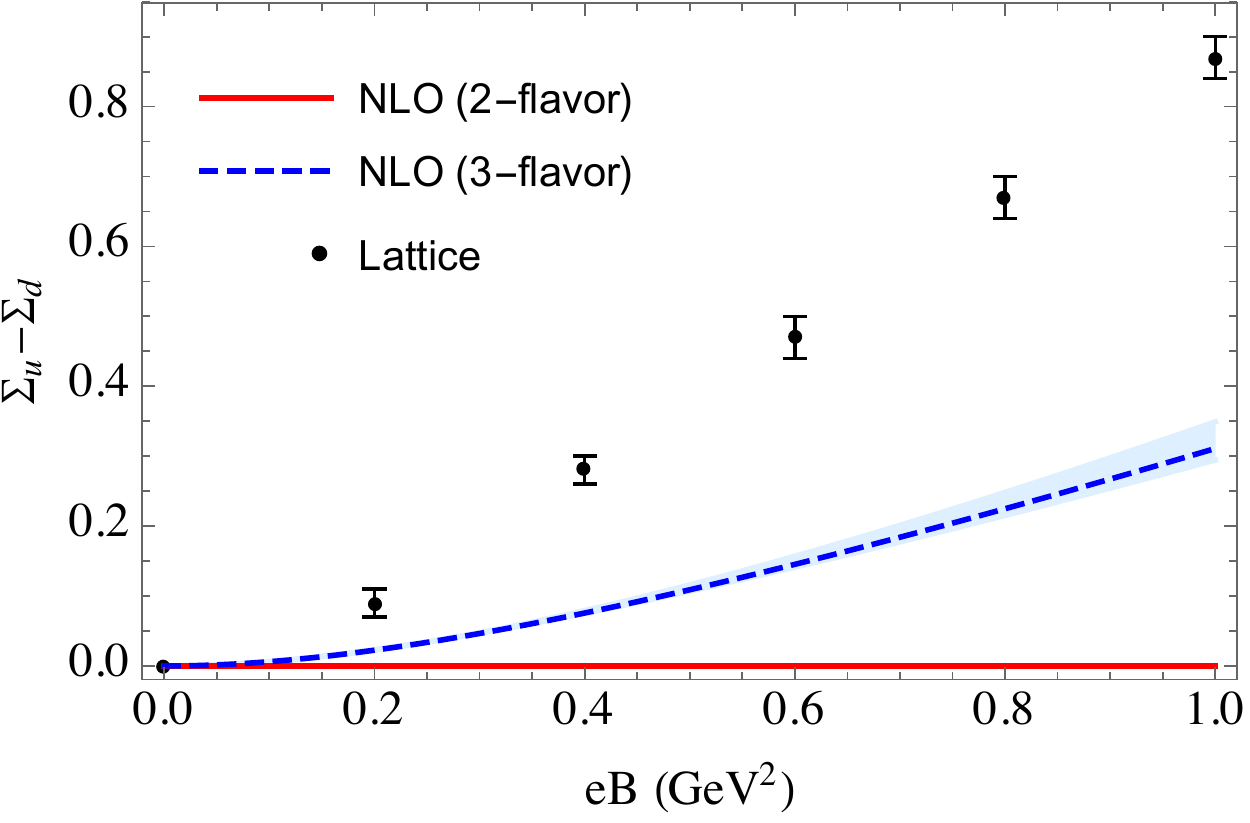}
  \caption{Plot the difference in the condensate shift, $\Sigma_{u}-\Sigma_{d}$, as a function of $eB$ at NLO in two-flavor $\chi$PT (red) and three-flavor $\chi$PT (blue). See main text for details.}
\label{conddiff}
\end{figure}

\begin{figure}[htb]
\centering  \includegraphics[width=0.45\textwidth]{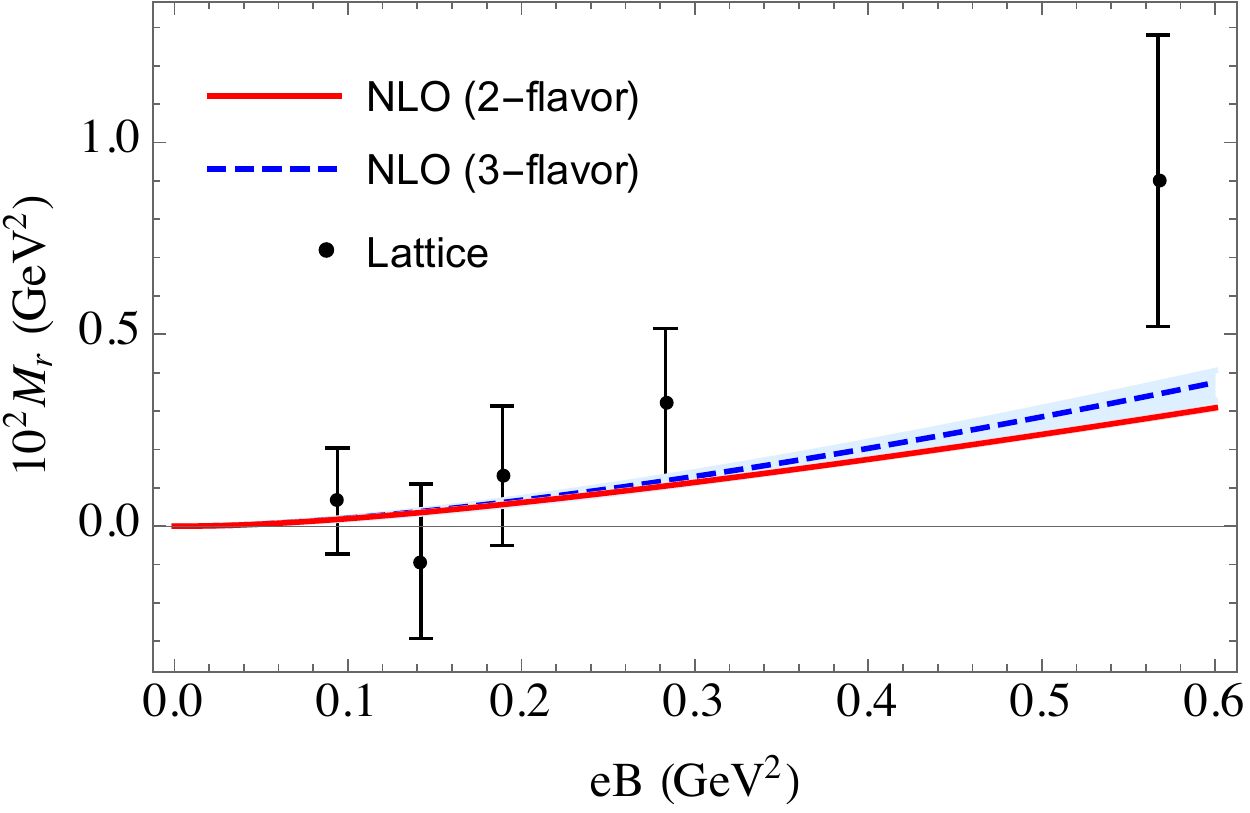}
  \caption{Plot of the renormalized magnetization, $M_{r}$, as a function of $eB$ at NLO in two-flavor $\chi$PT (red) and three-flavor $\chi$PT (blue). See main text for details.}
\label{magnetization}
\end{figure}

The shift in the light quark condensate is defined in the lattice study as~\cite{qqlat}
\bqa
\label{relshiftq}
\Sigma_{q}=-\frac{m}{m_{\pi}^{2}F^{2}}\langle\bar{q}q\rangle_{B}\ ,
\eqa
where $q$ refers to the quark flavor, $m$ is the average light quark mass, $\langle \bar{q}q\rangle_{B}$ is the shift in the chiral condensate due to the background magnetic field $B$, $m_{\pi}$ is the physical pion mass and $F$ is the pion decay constant in the chiral limit. The lattice uses the following parameters
\bqa
F&=&86\ {\rm MeV}\;,\\
m_{\pi}&=&135\ {\rm MeV}\;,\\
m_{K}&=&495\ {\rm MeV}\;.
\eqa
We do not need to specify the quark masses in order to calculate $\Sigma_{q}$ in $\chi$PT since the explicit factor of $m$ in Eq.~(\ref{relshiftq}) is cancelled by the one in the condensate shift, $\langle\bar{q}q\rangle_{B}$, which is proportional to $B_{0}$, where
\bqa
B_{0}=\frac{m_{\pi,0}^{2}}{2m}\ ,
\eqa
with $m_{\pi,0}$ being the bare pion mass and $m$ being the average light quark mass. 
However, we do need to specify the bare pion mass using the physical pion mass, which in two-flavor $\chi$PT is
\bqa
m_{\pi}^{2}=m_{\pi,0}^{2}\left[1-\frac{m_{\pi,0}^{2}}{2(4\pi f)^{2}}\bar{l}_{3} \right ]\; ,
\label{mpi2}
\eqa
while the physical  pion decay constant is 
\bqa
f_{\pi}=f\left[1+\frac{m_{\pi,0}^{2}}{(4\pi f)^{2}}\bar{l}_{4}
\right]\;.
\eqa
\begin{widetext}
Since the lattice condensate shifts, $\Sigma_{q}$, are specified in terms of the pion decay constant in the chiral limit, $F$, it is worth noting that in the two-flavor case, $f=F$. However, in the three-flavor case this is not the case: we proceed by noting the one-loop renormalized pion and kaon masses, and the pion decay constant valid in the isospin limit~\cite{gasser2},
\bqa\nonumber
  m_{\pi}^2&=&
  m_{\pi,0}^2\left[1
    -\left(8{L}_4^r+8{L}_5^r-16{L}_6^r-16{L}_8^r
      +{1\over2(4\pi)^2}\log{\Lambda^2\over m_{\pi,0}^2}\right)
    {m_{\pi,0}^2\over f^2}
-16({L}_4^r-2{L}_6^r){m_{K,0}^2\over f^2}\right.\\ &&\left.
  +{m_{\eta,0}^2\over6(4\pi)^2f^2}\log{\Lambda^2\over m_{\eta,0}^2}
  \right]\;,
  \label{mpi}\\ 
m_{K}^2&=&
m_{K,0}^2\left[1
  -8\left({L}_4^r-2{L}_6^r\right){m_{\pi,0}^2\over f^2}
%\right.\\ &&\left.
    -8(2L_4^r+{L}_5^r-4L_6^r-2{L}_8^r){m_{K,0}^2\over f^2}
  -{m_{\eta,0}^2\over3(4\pi)^2f^2}
  \log{\Lambda^2\over m_{\eta,0}^2}
\right]\;, 
\label{mk}
\\
f_{\pi}
&=&f\left[1
  +\left(4{L}_4^r
+4L_5^r+{1\over(4\pi)^2}\log{\Lambda^2\over m_{\pi,0}^2}
\right){m_{\pi,0}^2\over f^2}+\left(8L_4^r+{1\over2(4\pi)^2}\log{\Lambda^2\over m_{K,0}^2}
  \right){m_{K,0}^2\over f^2}
\right]\;,
\label{fpi}
\eqa
\noindent
where $m_{\pi,0}$ and $m_{K,0}$ are the bare and degenerate three-flavor pion and kaon masses respectively, and $f$ is the bare pion (or kaon) decay constant~\cite{gasser2}. In the chiral limit, Eq.~(\ref{fpi}) reduces to
\end{widetext}
\bqa
\label{fpi0}
F
&=&f\left[1+\left(8L_4^r+{1\over2(4\pi)^2}\log{\Lambda^2\over m_{K,0}^2}
  \right){m_{K,0}^2\over f^2}
\right]\ ,
\eqa
where we note that unlike two-flavor $\chi$PT, $f\neq F$. 
For our comparison, the only two-flavor LEC
needed is~\cite{bijnensreview}
\begin{equation}
\begin{split}
\bar{l}_{3}=2.9\pm 2.4\ ,
\end{split}
\end{equation}
%{\color{green}evaluated at which scale $m_{\pi,0}$}. 
Using Eq.~(\ref{mpi2}), gives the following bare pion mass with uncertainties
\begin{equation}
\begin{split}
m_{\pi,0}=136.6^{+1.4}_{-1.3}\ {\rm MeV}\ \textrm{(two-flavor)} \ .
\end{split}
\end{equation} 
We also require the three-flavor couplings $L_i^r$; since they are scale-dependent, they are specified at a certain scale.
In Ref.\cite{bijnensreview}, the scale
is $\mu=0.77\ {\rm GeV}$ 
which is approximately  the mass of the
$\rho$ meson and where
$\Lambda^2=4\pi e^{-\gamma_E}\mu^2$,
\begin{align}
{L}_{4}^r&=(0.0 \pm 0.3)\times10^{-3}\;,\\
{L}_{5}^r&=(1.2 \pm 0.1)\times10^{-3}\;,\\
{L}_{6}^r&=(0.0 \pm 0.4)\times10^{-3}\;,\\
{L}_{8}^r&=(0.5 \pm 0.2)\times10^{-3}\;.
\label{LECs}
\end{align}
We then use Eqs. (\ref{mpi}), (\ref{mk}), and (\ref{fpi0}) to calculate the bare parameters,
\begin{align}
f&=93.4^{+11.2}_{-7.4}{\rm \ MeV}\ \textrm{(three-flavor)}\;,\\
m_{\pi,0}&=135.1^{+18.3}_{-9.1}{\rm\ MeV}\ \textrm{(three-flavor)}
\;,\\
m_{K,0}&=474.1^{+76.3}_{-47.7}{\rm \ MeV}\ \textrm{(three-flavor)}
\;,
\end{align}
with the uncertainties giving rise to the bands in the plots of the condensate shifts in Figs.~\ref{condavg}, \ref{conddiff} and \ref{magnetization}. We plot the two-flavor, one-loop results using red, solid lines with light red bands indicating uncertainties due to the two-flavor LECs. Similarly, one-loop results from three-flavor $\chi$PT are indicated using dashed, blue lines with light band shades indicating the uncertainties due to the three-flavor LECs.

In Fig.~\ref{condavg} we plot the average of the up and down condensate shifts, ${1\over2}(\Sigma_{u}+\Sigma_{d})$. We note that the one-loop three-flavor result (blue, dashed) is a significant improvement over the one-loop, two-flavor result, particularly for larger values of $eB$. However, neither the two-flavor nor the three-flavor, one-loop results agree very well with the lattice results with the three-flavor results in modest and improved agreement for values of $eB<0.3\ {\rm GeV^{2}}$ compared to two-flavor results.
The uncertainties in the LECs translate into a
large uncertainty band in the three-flavor case, which does not preclude the possibility of the average light quark condensate being smaller than in the the two-flavor case.

In Fig.~\ref{conddiff}, we plot the difference in the up and down quark condensate, $\Sigma_{u}-\Sigma_{d}$. We note that away from the isospin limit, the up and down quark condensates are different in the absence of a magnetic field in two-flavor $\chi$PT~\cite{gasser1}, the shift due to a magnetic field is somewhat surprisingly independent in two-flavor $\chi$PT in spite of the difference in the charges of up and down quarks~\cite{delia}. The physical reason for this lies in the fact that the charged degrees of freedom in two-flavor $\chi$PT interact in the same manner with the external magnetic field modulo the difference in the sign of their electromagnetic charges. However, in three-flavor $\chi$PT the shift is positive since the charged degrees of freedom are the pions and kaons with both containing up valence quarks, which have a larger magnitude of charge compared to the down and strange quarks. Furthermore, at one-loop order in three-flavor $\chi$PT, the three quark condensate shifts are related, $\langle \bar{u}u\rangle_{B}=\langle \bar{d}d\rangle_{B}+\langle \bar{s}s\rangle_{B}$. As such the condensate shift difference can be interpreted as the magnitude of the shift in the strange quark condensate modulo an overall factor of $B_{0}$. However, there is currently no lattice data currently available for a direct comparison.

Finally in Fig.~\ref{magnetization}, we compare the renormalized magnetization (density) of the QCD vacuum in a magnetic field from $\chi$PT that of lattice QCD. Such a comparison has been previously done in the context of the Hadron Resonance Gas Model~\cite{hrg}. In the two-flavor case, only the pions contribute to the renormalized magnetization while in the three-flavor case there is a contribution from the charged pions as well as the charged mesons. As Fig.~\ref{magnetization} shows, the renormalized magnetization increases due to the additional contribution of the charged kaons, including uncertainties. We observe that the increase in the magnetization is more prominent for larger values of $eB$, i.e. the vacuum becomes more paramagnetic. Furthermore, the three-flavor magnetization is in slightly better agreement with the magnetization from the lattice than the two-flavor magnetization though the uncertainties in the lattice data are large.

Finally, we note that it is of interest to include two-loop corrections to the 
average light condensate shift, the difference in the condensate shift, and the magnetization, in light of the
fact that these quantities are underestimated relative to lattice result. 
Work in this direction is currently in progress~\cite{3fus}.
%\onecolumngrid
%\newpage

\section*{Acknowledgements}
The authors would like to acknowledge Nikihea Agasian and Brian Tiburzi for useful discussions. We would also like to acknowledge Gergo Endr\H{o}di for sharing lattice data and helpful discussions. P.A. would also like to acknowledge Wellesley College, where the initial stages of this work was done and Saint Olaf College for startup funds.

\appendix
\section{Integrals}
\label{app:integrals}
We use dimensional regularization to regulate the integrals.
They are defined in $d=4-2\epsilon$ dimensions
\bqa
\label{intdef}
\int_p&=&\left({e^{\gamma_E}\Lambda^2\over4\pi}\right)^{\epsilon}
\int{d^dp\over(2\pi)^d}\;,
\eqa
where $\Lambda$ is the renormalization scale associated with the
$\overline{\rm MS}$ scheme. For charged mesons, the integrals are
over $p_{\parallel}$ with $p_{\parallel}^2=p_0^2+p_z^2$,
defined in $d=2-2\epsilon$ dimensions .
The integrals we need are
\bqa\nonumber
I_n^B(m^2)&=&-{\left(e^{\gamma_E}\Lambda^2\right)^{\epsilon}\over(4\pi)^2}
\int_0^{\infty}{ds\over s^{4-n-\epsilon}}e^{-m^2s}{eBs\over\sinh(eBs)}\\
&=& \nonumber
-{e^{\gamma_E\epsilon}\over(4\pi)^2}\left(2|qB|\right)^{3-n}
\left({\Lambda^2\over2eB}\right)^{\epsilon}
\\ && \times
\Gamma[n-2+\epsilon] 
\zeta(n-2+\epsilon,x+\mbox{$1\over2$})\;,
\label{indef}
\eqa
where $x={m^2\over2eB}$, $\Gamma[x]$ is the Gamma function, and 
$\zeta(a,x+{1\over2})$ is the Hurwitz zeta function~\cite{grad}.
The integrals $I_n^B(m^2)$ are well defined in the chiral limit.
For $\epsilon=0$, the integrals are ultraviolet divergent for $n\leq3$.
The integrals $I_n^B$ satisfy the recursion relation
\bqa
{dI_n^B\over dm^2}&=&-I_{n-1}^B\;.
\eqa
The $B$-dependence of the  integrals $I_n$ can be isolated 
so that we write them $I_n^B(m^2)=I_n(m^2)+\tilde{I}_n^B(m^2)$.
Specifically, we need $I_1^B(m^2)$ and $I_2^B(m^2)$. Their expansion in powers of $\epsilon$ is
\bqa
I_1(m^2)&=&-{m^{4}\over2(4\pi)^2}\left[{1\over\epsilon}+{3\over2}+\log{\Lambda^2\over m^2}
+{\cal O}(\epsilon)
\right]\;,
\\ \nonumber
    \tilde{I}_{1}^{B}(m^{2})&=&\frac{(eB)^{2}}{6(4\pi)^{2}}\left[\frac{1}{\epsilon}+\log\frac{\Lambda^{2}}{m^{2}}+\mathcal{O}(\epsilon)\right]\\
    &&+\frac{(eB)^{2}}{(4\pi)^{2}}\delta\tilde{I}_{1}^{B}(m^{2})\;,
    \\ \nonumber
    &&\\
\delta\tilde{I}_{1}^{B}(m^{2})
 &=& \nonumber
\bigg[4\zeta^{(1,0)}(-1,x+\tfrac{1}{2})+x^{2}
 \bigg. \\
    &&\left.-2x^{2}\log x+\frac{1}{6}\log x+\frac{1}{6}\right]\;,
    \label{dI1B}
\\
I_2(m^2)&=&
{m^2\over(4\pi)^2}\left[{1\over\epsilon}+1+\log{\Lambda^2\over m^2}
+{\cal O}(\epsilon)
\right]\;,\\ \nonumber
\tilde{I}_2^B(m^2)&=&-{2eB\over(4\pi)^2}
\left[
\zeta^{(1,0)}(0,x+\tfrac{1}{2})+x-x\log x\right]\;.
\\ &&
\label{i2b}
\eqa

%\bibliographystyle{apsrev4-1}

%\bibliography{/Users/prabal7e7/Documents/Research/bib}
%\bibliography{bib
\end{document}